# Pay or Play

**Sigal Oren** *  
Cornell University

**Michael Schapira**  
Hebrew University  
and Microsoft Research

**Moshe Tennenholtz**  
Technion - Israel Institute of Technology  
and Microsoft Research

## Abstract

We introduce the class of *pay or play* games, which captures scenarios in which each decision maker is faced with a choice between two actions: one with a fixed payoff and another with a payoff dependent on others' selected actions. This is, arguably, the simplest setting that models selection among certain and uncertain outcomes in a multi-agent system. We study the properties of equilibria in such games from both a game-theoretic perspective and a computational perspective. Our main positive result establishes the existence of a semi-strong equilibrium in every such game. We show that although simple, pay or play games contain well-studied environments, e.g., vaccination games. We discuss the interesting implications of our results for these environments.

## 1 Introduction

The situation in which a decision-maker has to choose between an action with fixed, certain, outcome to a course of action with uncertain consequences is a fundamental topic in decision making under uncertainty. We introduce a new framework, called *pay or play*. In pay or play each of multiple decision makers must choose among an action with a known, fixed, payoff, and an action interpreted as participation in a game with other decision makers. The outcome of this game is dependent on who of the other decision makers also choose to take part in this game. The pay or play setting captures what is arguably the simplest scenario in which decision makers select between certain and uncertain outcomes, and the realization of the uncertain outcome is solely dependent on the decision makers and not on "nature". Importantly, in addition to its theoretical and conceptual appeal, pay or play encompasses, unifies, and abstracts classical models of immunization and of differential pricing.

**A Game-Theoretic Formulation.** We now give an informal, high-level, exposition of our (game-theoretic) pay or play model. In a pay or play game there are $n$ self-interested players, each with two possible strategies (actions). Each player $i$ has a cost function $c_i$ which specifies, for every $n$-tuple of players' strategies, the cost of player $i$. $c_i$ is such that whenever player $i$'s strategy is pay his cost is some fixed value $h_i$, regardless of what the other players' strategies are. When player $i$'s strategy is play, however, his cost is a function of the other players whose strategy is also play. We require each cost function $c_i$ to be monotone nondecreasing, i.e., as more players choose play the cost of player $i$ cannot decrease.

We are interested in the properties of (Nash) equilibria in this game-theoretic setting. An equilibrium is an $n$-tuple of strategies from which no player wishes to unilaterally deviate. We explore both pure (deterministic) equilibria, in which each player must choose one of these two strategies, and mixed (randomized) equilibria in which a player can choose a probability distribution over the two strategies. We tackle fundamental questions, including: Does a pure equilibrium always exist? Are equilibria in this environment "globally efficient"? What is the complexity of determining the existence and computing equilibria? And more.

**Our Contributions.** We study the properties of equilibria in pay or play games both from a game-theoretic perspective and a computational perspective. We now briefly summarize our results:

We begin by showing that a pure Nash equilibrium may not always exists and characterize some subclasses pay or play games which always admit a pure Nash equilibrium. The next natural question is how

---

hard is it to determine whether such an equilibrium exists or not—a question tackled in a large variety of other game-theoretic contexts. We show that this task is, in general, intractable from both a computational perspective (NP-hard) and information-theoretic (communication complexity) perspective.

A main criticism against Nash equilibria is that they are not resilient to deviations by coalitions of players. Equilibria that are resilient against all such deviations, called "strong equilibria", are hence of special interest. We identify conditions for the existence of a strong equilibrium. Our main positive result is that *any* pay or play game admits an equilibrium with a slightly weaker property, namely, a "semi-strong" equilibrium.

Next, we explore the conditions under which pay or play games are Pareto efficient, i.e., when no scenario that is strictly better for at least a single player and no worse for all others exists. We also quantify the gap in global efficiency (sum of players' costs) between an equilibrium and the optimum solution (which does not take into account players' own selfish agendas).

Lastly, we discuss the implications of our results for two special cases of pay or play games: classical models of immunization [1, 2] and of differential pricing [15, 16]. In particular, we show that the game described in [1] always admits a Pareto efficient pure Nash equilibrium.

**Related Work** Decision between actions with certain and uncertain outcomes is the subject of much research in decision theory. Indeed, the rich literature about the (so called) value of information, which concentrates on measuring the gain one obtains by acquiring information. See, e.g., [6, 12, 5, 13].

Equilibrium analysis is fundamental to game theory and has recently also received much attention from a computer science perspective. In particular, establishing when different kinds of equilibria (pure Nash equilibrium, strong Nash equilibrium, and more) are guaranteed to exist, and the complexity of computing such equilibria, are two important, and extensively studied, research topics. See, e.g., classical game-theoretic results on the existence of pure Nash equilibria in congestion games [14], potential games [11], and player-specific congestion games [10], and also more recent results on computing equilibria in these environments [3, 4].

We have already mentioned that pay or play games generalizes classical models of immunization and differential pricing. An additional class of games generalized by the pay or play class are Interdependent Security Games [7, 8]. Similarly to immunization games in these games players decide whether to invest in security or not and their decision affects both their own vulnerability and their peers vulnerability.

## 2 Model and Preliminaries

In pay or play games we have a set of $N$ self-interested players ($|N| = n$), each with two strategies: *pay* or *play*. We denote the choices of the players by a strategy vector $x = (x_1, \ldots, x_n)$. When referring to pure (deterministic) strategy profiles, that is, the scenario that each player selects either pay or play with probability 1, we shall use $x_i = 0$ to indicate that player $i$ chooses the play strategy and $x_i = 1$ to indicate that player $i$ chooses the pay strategy. We denote by $A(x)$ the set of players who choose the play strategy in pure strategy vector $x$. The cost of player $i$ in pure strategy-vector $x$, $c_i(x)$, is some fixed number $h_i$ if $i$ pays in $x$ (i.e., $x_i = 1$) and a function of the set of players who play in $x$, $g_i(A(x))$, if $i$ plays in $x$. Formally, we define:

$$c_i(x) = \begin{cases} h_i & x_i = 1 \\ g_i(A(x)) & x_i = 0 \end{cases}$$

In cases that all the players have the same cost function we will refer to the fixed cost as simply $h$ and the cost of the play strategy as $g(\cdot)$.

We require $g_i(\cdot)$ to be monotone nondecreasing (that is, as more players choose play the cost of player $i$ should increase). Formally, if $S \subseteq T$ and $i \in S$ then $g_i(S) \leq g_i(T)$.

Recall that a player plays a mixed strategy when he selects some probability distribution over the two actions. For mixed strategies, $x_i$ will denote the probability that player $i$ chooses the pay strategy. (Observe that a pure strategy is a special case of a mixed strategy.) The cost of player $i$ in a mixed strategy vector $x$, $c_i(x)$, is his expected cost over the induced distribution over pure strategy vectors: $c_i(x) =$

$$x_i \cdot h_i + (1 - x_i) \cdot \sum_{S \subseteq N - \{i\}} \prod_{j \in S} (1 - x_j) g_i(S \cup \{i\}).$$

Our focus in this paper is on the Nash equilibria of play or play games that are defined as follows:

**Definition 2.1** *A vector of mixed (pure) strategies $x$ is a* m*ixed (pure) Nash equilibrium if for every player $i$ and every mixed (pure) strategy $x_i'$: $c_i(x_i', x_{-i}) \geq c_i(x)$.*

As common is game theory literature, $x_{-i}$ is used as shorthand for the strategy vector describing all players' strategies but that of player $i$, and $(x_i, x_{-i})$ denotes the strategy vector in which player $i$'s strategy is $x_i$ and the other players' strategies are as in $x_{-i}$.

# 3 Pure Nash Equilibria

We begin by addressing the natural question of whether a pure Nash equilibrium always exists in pay or play games. We provide an affirmative answer to this question for some subclasses of pay or play games, but show that, in general, the class includes games that do not admit a pure Nash equilibrium. Furthermore, we show that determining whether a specific pay or play game admits a pure Nash equilibrium is hard both from a computational perspective (NP-hardness) and from an information-theoretic perspective (can involve the communication of exponentially many bits). As each player has a two strategies, though not all pay or play games possess a pure Nash equilibrium, all games do admit at least a single mixed Nash equilibrium. We discuss the properties of such equilibria later on.

## 3.1 Sufficient Conditions for Existence

Note that pay or play games in which (i) the cost functions of all players depend only on the number of players who choose the play strategy *and* (ii) all players have the same cost function belong to the classic game-theoretic category of "congestion games" [14], and so are guaranteed to possess a pure Nash equilibrium. We now show that a sufficient condition for a play or pay game to admit a pure Nash equilibrium is for just one of these two properties to hold.

First, consider pay or play games in which the cost function of the play strategy ($g_i(\cdot)$) of all the players depends only on the number of players who choose the play strategy (and not on their identities). It is not hard to observe that such games belong to the class of *player-specific congestion games*. This class of games was defined by Milchtaich [10], who showed that these games always admit a pure Nash equilibrium. Thus, the following claim holds for pay or play games:

**Claim 3.1** *If for every player $i$ there exists a function $w_i$ such that for every $S \subseteq N - \{i\}$, $g_i(S \cup \{i\}) = w_i(|S|+1)$, then a pure Nash equilibrium exists.*

We now show that if the players are symmetric (i.e., all have the same cost function), then a pure Nash equilibrium always exists. We point out that the cost function of the players is allowed to depend on the identities of players who choose to play (and not just on their number).

**Claim 3.2** *If all players in a pay or play game are symmetric, then a pure Nash equilibrium of the game always exists and can be computed efficiently.*

**Proof:** We present a simple greedy algorithm for computing a pure Nash equilibrium in polynomial time: begin with the strategy vector $x = 1^n$ in which all players choose the pay strategy. While there exists a player $i \notin A(x)$ such that $g(A(x) \cup \{i\}) < h$ set $x_i = 0$.

We claim that the resulting strategy vector is a pure Nash equilibrium. Observe that once the algorithm halts every player $i \in A(x)$ has a cost smaller than $h$, and so prefers the play strategy. On the other hand, every player $j \notin A(x)$ would have a cost greater than $h$ for choosing the play strategy. □

## 3.2 Computational Hardness

Next, we show that if the costs are both player-specific and can depend on the identities of the players, a pure Nash equilibrium might not exist at all. This is true even when the cost functions are restricted to be submodular [1].

**Claim 3.3** *The pay or play class contains games that do not admit a pure Nash equilibrium, even for submodular cost functions.*

**Proof:** Consider the following game consisting of three players numbered $0, 1, 2$. The cost of player $i$ is defined as: $h_i = 1.5$, $g_i(\{i-1, i, i+1\}) = 2$, $g_i(\{i-1, i\}) = 2$, $g_i(\{i, i+1\}) = 1$, $g_i(\{i\}) = 1$. Where $i+1$ and $i-1$ are computed modulo 3. We show that this game does not admit any pure Nash equilibrium by doing a case by case analysis of all the possible strategy vectors:

- There is no pure Nash equilibrium in which all players choose the play strategy – one of the players can benefit from choosing the pay strategy.
- There is no pure Nash equilibrium in which two players choose the play strategy – if players $j$ and $j+1$ choose the play strategy then the cost of player $j+1$ is 2 and hence he prefers to choose the pay strategy.
- There is no pure Nash equilibrium in which at most a single player chooses the play strategy – if players $j$ and $j+1$ choose the pay strategy then player $j$ can reduce his cost to 1 by switching to the play strategy.

□

We are now ready to show that determining whether a pure Nash equilibrium exists or not is NP-hard. The proof is based on a reduction from a 3-SAT formula to a pay or play game and uses the construction from the previous claim as a gadget.

---

[1] A cost function $g(\cdot)$ is submodular if for every two sets of players $S \subseteq T$ and for ever player $j \notin T$ it holds that: $g(T \cup \{j\}) - g(T) \leq g(S \cup \{j\}) - g(S)$.

**Theorem 3.4** *Determining whether a pure Nash equilibrium exists or not in a pay or play game is NP-hard.*

**Proof:** Given an instance of 3-SAT we construct the following pay or play instance where all players have the same fixed cost of 1.5 but different cost functions for the play strategy.

- For each variable $v_i$ of the 3-SAT formula, we create two players – $t_i$ and $f_i$. We construct their cost functions such that whenever $f_i$ chooses to play then $t_i$ prefers to pay and vice versa. Formally, we define for all subsets $S$ such that $f_i \in S$: $g_{t_i}(S) = 2$ and for all $S$ such that $f_i \notin S$: $g_{t_i}(S) = 1$. Similarly, we define for all $S$ such that $t_i \in S$: $g_{f_i}(S) = 2$ and for all $S$ such that $t_i \notin S$: $g_{f_i}(S) = 1$.
- For every clause $i$ we create three players, $a_{3i}, a_{3i+1}, a_{3i+2}$. We find it easiest to define their costs by an example: consider, for instance, $i = (v_j \vee \bar{v}_k \vee v_l)$, if $t_j \notin S$ or $f_k \notin S$ or $t_l \notin S$ then $g_{a_{3i+r}}(S) = 1$ for $r \in \{0,1,2\}$. Else, for a set $S$ such that $t_j, f_k, t_l \in S$ and $a_{3i}, a_{3i+1}, a_{3i+2} \notin S$, we reconstruct the example from Claim 3.3 and define:
  - $g_{3i+r}(\{a_{3i+r-1}, a_{3i+r}, a_{3i+r+1}\} \cup S) = 2$
  - $g_{3i+r}(\{a_{3i+r-1}, a_{3i+r}\} \cup S) = 2$
  - $g_{3i+r}(\{a_{3i+r}, a_{3i+r+1}\} \cup S) = 1$
  - $g_{3i+r}(\{a_{3i+r}\} \cup S) = 1$

  where $r+1$ and $r-1$ are computed modulo 3.

**Claim 3.5** *The 3-SAT formula can be satisfied if and only if the previously defined game admits a pure Nash equilibrium.*

**Proof:** First assume that the formula is satisfiable. Let $\phi$ be an assignment satisfying it. We show that the following strategy vector is an equilibrium: every player of type $a_i$ uses the play strategy, player $t_i$ chooses the play strategy if and only if $\phi_i = T$ and player $f_i$ chooses the play strategy if and only if $\phi_i = F$. To verify that this is indeed a Nash equilibrium observe the following: first, for every $i$ player $a_i$ has a cost of 1 which is smaller than the cost of 1.5 for choosing the pay strategy. If player $t_i$ uses the pay strategy, then player $f_i$ does not use the pay strategy – thus the cost of player $t_i$ for using the pay strategy is 1.5, if it instead chooses the play strategy it would pay 2. Player $f_i$ cost is 1 for playing so this is its best response as well. Similarly, one can show that this is also an equilibrium for players $t_i$ and $f_i$ such that $t_i$ uses the play strategy and player $f_i$ uses the pay strategy.

Next, we show that if there exists a pure Nash equilibrium then the formula is satisfiable. Let $x$ be the Nash equilibrium. Clearly it has to be the case that for all pairs $f_i, t_i$ exactly one of the players chooses pay and the other chooses play. Consider the assignment $\phi_i = T$ if $x_{t_i} = 1$ and $\phi_i = F$ if $x_{t_i} = 0$. Assume towards a contradiction that there exists some clause $i$ which is not satisfied by the assignment $\phi$. Suppose, for instance, that $i = (v_j \vee \bar{v}_k \vee v_l)$. This implies that, $t_j, f_k$ and $t_l$ all use the play strategy. Therefore, by construction the three players $a_{3i}, a_{3i+1}, a_{3i+2}$ are in the exact same configuration as the nodes in Claim 3.3 and thus a Nash equilibrium does not exist. □ □

## 3.3 Communication Hardness

We now prove that determining whether a pure Nash equilibrium exists in a pay or play game is also hard from an information-theoretic perspective. Specifically, we consider the problem of determining whether a Nash equilibrium exists in Yao's classic communication complexity model [17]: Suppose that each of the $n$ players in a pay or play game knows only his own cost function and the different players wish to find out whether, when put together, their cost functions induce a game that admits a pure Nash equilibrium. No computational restrictions whatsoever are imposed on the players. We set an exponential (in the number of players, $n$) lower bound on the number of bits the players must exchange to learn the answer to this question. (Observe that a player cannot always simply reveal his entire cost function to others as its specification can, in general, be exponential in $n$.)

**Theorem 3.6** *Determining whether a Nash equilibrium exists in a pay or play game requires communicating an exponential (in n) number of bits.*

**Proof:** To prove the lower bound we present a reduction from the well-studied DISJOINTNESS problem from communication complexity theory. In this classical setting, there are two parties 1 and 2, each holding a subset $A_i \subseteq \{1, \ldots, r\}$. The objective in DISJOINTNESS is to distinguish between the following two possibilities: (1) $A_1 \cap A_2 \neq \emptyset$ (2) $A_1 \cap A_2 = \emptyset$.

Classical results in communication complexity establish that solving DISJOINTNESS necessitates (in the worst case) transmitting $\Omega(r)$ bits. For more information the interested reader is referred to [9].

We now show how to construct an $n$-player pay or play game $G$ such that a pure Nash equilibrium in $G$ exists if and only if $A_1 \cap A_2 \neq \emptyset$ in the DISJOINTNESS instance. Suppose that $r = \binom{\frac{n-6}{2}}{\frac{n-6}{4}}$ (w.l.o.g., let $n = 4k+6$ for some integer $k > 0$). We identify each element $j \in \{1, \ldots, r\}$ with a unique set $S_j \subseteq \{1, \ldots, \frac{n-6}{2}\}$ of size $\frac{n-6}{4}$. We create $n-6$ players as follows. For every element $j \in \{1, \ldots, \frac{n-6}{2}\}$ we create two players $v_j$ and

$u_j$. We construct their cost functions such that whenever $v_j$ chooses to play $u_j$ prefers to pay and vice versa. Formally, $v_j$'s cost when choosing the pay strategy is 1.5, as for the play strategy, for all subsets of players $S$ such that $u_j \in S$: $g_{v_j}(S) = 2$, and for all $S$ such that $u_j \notin S$: $g_{v_j}(S) = 1$. The cost function of player $u_j$ is defined similarly.

We create 6 more players: $t_0, t_1, t_2$, and $w_0, w_1, w_2$. The cost functions of each of the three players $t_0, t_1$, and $t_2$ are similar to those in the example from Claim 3.3 and are defined as follows: the cost of player $t_i$, $h_{t_i} = 1.5$; for any set $S \subseteq \{1, ..., \frac{n-6}{2}\}$ let $V_S = \bigcup_{i \in S}\{v_i\}$; if there is some $j \in A_1$ such that $S_j \subseteq S$, $g_{t_i}(V_S) = 2$; if $S_j$ is not contained in $S$ for any $j \in A_1$, $g_{t_i}(t_{i-1}, t_i, t_{i+1}, V_S) = 2$, $g_{t_i}(t_{i-1}, t_i, V_S) = 2$, $g_{t_i}(t_i, t_{i+1}, V_S) = 1$, $g_{t_i}(t_i, V_S) = 1$, where $i+1$ and $i-1$ are computed modulo 3. The cost functions of each of the three players $w_0, w_1, w_2$ are defined similarly: the cost of player $w_i$, $h_{w_i} = 1.5$; for any set $S \subseteq \{1, ..., \frac{n-6}{2}\}$ let $U_S = \bigcup_{i \in S}\{u_i\}$; if there is some $j \in A_2$ such that $S_j^C \subseteq S$, where $S_j^C$ denotes the complement of $S_j$, then $g_{w_i}(U_S) = 2$; otherwise, $g_{w_i}(w_{i-1}, w_i, w_{i+1}, U_S) = 2$, $g_{w_i}(w_{i-1}, w_i, U_S) = 2$, $g_{w_i}(w_i, w_{i+1}, U_S) = 1$, $g_{w_i}(w_i, U_S) = 1$. $i+1$ and $i-1$ are again computed modulo 3.

**Claim 3.7** *There is a Nash equilibrium in the pay or play game $G$ if and only if $A_1 \cap A_2 \neq \emptyset$ in the DISJOINTNESS instance.*

**Proof:** First consider the scenario that $A_1 \cap A_2 \neq \emptyset$ in the original DISJOINTNESS instance. We show that in this case there is indeed a pure Nash equilibrium in $G$. Let $j \in A_1 \cap A_2$. For every $i \in S_j$ set the strategy of player $v_i$ to be play and the strategy of player $u_i$ to be pay. For every $i \in \{1, ..., \frac{n-6}{2}\} \setminus S_j$ set the strategy of player $v_i$ to be pay and the strategy of player $u_i$ to be play. Observe that none of the $v_i$'s or $u_i$'s wish to unilaterally deviate from this (still partial) specification of players' strategies as each of these players' strategies is the exact opposite of that of his counterpart. Now, set the strategies of all $t_i$'s and $w_i$'s to be pay. Observe the $t_i$'s do not wish to deviate as the set of $v_i$-players who chose to play corresponds to the set $S_j$. Observe also that the $w_i$'s do not wish to deviate as the set of $u_i$'s who chose to play corresponds to the set $S_j^C$.

Next, we show that if there exists a Nash equilibrium then $A_1 \cap A_2 \neq \emptyset$. We make the following crucial observation: in any Nash equilibrium exactly $\frac{n-6}{4}$ of the $v_i$'s are using the play strategy. To see this, consider a specific Nash equilibrium. Observe that if more than $\frac{n-6}{4}$ $v_i$'s choose to play then in any pure Nash equilibrium their $u_i$ counterparts would choose to pay. This means that less than $\frac{n-6}{4}$ $u_i$'s pay, which in turn means that, by construction, the three players $w_0, w_1, w_2$ are in the exact same configuration as the nodes in Claim 3.3—this leads to a contradiction, since for the three nodes in this configuration a pure Nash equilibrium does not exist. A similar argument establishes that no less than $\frac{n-6}{4}$ of the $v_i$'s must play in any Nash equilibrium as otherwise the $t_i$'s will find themselves in the same predicament. Consider now the case that exactly $\frac{n-6}{4}$ $v_i$'s play. Observe that the $t_i$'s avoid being in the configuration in Claim 3.3 only if the set of $v_i$'s who play corresponds to some $S_j$ where $j \in A_1$ and the same holds for the $w_i$ players only if the set of $u_i$ who chose play corresponds to $S_j^C$ and $j \in A_2$. Hence, $j \in A_1 \cap A_2$. □ □

## 4 Strong and Semi-Strong Equilibria

One of the criticism often raised against Nash equilibria is that they are not resilient to deviations by coalitions of players. Hence, games that admit an equilibrium that is resilient against deviations by coalitions are of special interest. Such equilibria are called *strong* equilibria.

**Definition 4.1** *An equilibrium $x$ is strong if there is no strategy vector $y$, such that, for every player $i \in \{j | x_j \neq y_j\}, c_i(y) < c_i(x)$. When $y$ is restricted to be a pure strategy vector we say that $x$ is strong with respect to pure deviations.*

We show that pay or play games that admit a pure Nash equilibrium also admit a strong pure Nash equilibrium:

**Theorem 4.2** *If there exists a pure Nash equilibrium in a pay or play game then this equilibrium is strong with respect to pure deviations.*

**Proof:** Let $x$ be a pure Nash equilibrium. Assume towards contradiction that there exists a deviation of the set of players $S$ that reduces the cost of all of them. Observe that $S$ cannot include any player $i$ that previously used the play strategy ($x_i = 0$). The cost of such players is at most $h_i$ since $x$ is an equilibrium and by switching to the pay strategy their cost would be exactly $h_i$. Thus, the set consists of players that use the pay strategy in $x$ ($x_i = 1$) and deviate to the play strategy. However, by monotonicity, if player $i$ prefers the play strategy when more players are choosing it, then he should also prefer it when a smaller subset is playing it – in contradiction to the fact that $x$ is an equilibrium. □

One might also require the stronger property that an equilibrium would be also resilient against (uncoordinated) mixed deviations. Unfortunately, as the follow-ing example demonstrates, Nash equilibria (both pure and mixed) in our games are not necessarily strong with respect to mixed deviations.

**Example 4.3** *Consider the following symmetric 2-player instance: the cost of the pay strategy is $2+\epsilon$, for some small $\epsilon$. The cost of the play strategy is $2$ if both players choose it and $1$ if only one of them chooses it. The* unique *equilibrium is for both players to choose the play strategy. Observe that this equilibrium is not resilient against mixed deviations: if the two players choose the play strategy each one exhibits a cost of $2$. On the other hand, if they both deviate and use the mixed strategy of choosing to pay with probability $1/2$ to play with probability $1/2$, their cost is reduced to $\frac{1}{2}(2+\epsilon) + \frac{1}{2}(\frac{1}{2} \cdot 1 + \frac{1}{2} \cdot 2) = \frac{7}{4} + \frac{1}{2}\epsilon$.*

On the bright side, as we shall show below, the equilibria of games in our class are resilient against mixed deviations in a slightly weaker sense, called *semi-strong* Nash equilibrium. Roughly speaking, even though players can benefit from a joint deviation, this deviation is not "stable", as there always exists a player who can improve his cost by deviating again. For instance, the players in Example 4.3 could profit from jointly deviating to the mixed strategy $x_i = \frac{1}{2}$. However, after this deviation, each one of the players can decrease his cost even more by deviating to the strategy $x_i = 0$. The fact that deviations are not stable renders coalition formation hard (as there will always be a player who has an incentive to "betray" the others and deviate from the plan).

**Definition 4.4** *A mixed equilibrium $x$ is* semi-strong *if for every mixed strategy vector $y$ at least one of the following properties hold:*

1. *There exists a player $i$ such that $x_i \neq y_i$ and $c_i(y) > c_i(x)$.*
2. *There exists a player $i$ such that $x_i \neq y_i$ and a strategy $z_i \neq y_i$ such that $c_i(z_i, y_{-i}) < c_i(y)$.*

We are now ready to prove our main positive result: every equilibrium of a pay or play game is semi-strong. The proof is based on the following simple, yet powerful, fact: if player $i$ plays a mixed strategy then his expected cost is exactly $h_i$, since in a mixed equilibrium the player's two strategies should give the same payoff.

**Theorem 4.5** *Every mixed Nash equilibrium in a pay or play game is semi-strong.*

**Proof:** Consider an equilibrium $x$, assume towards a contradiction that it is not a semi-strong equilibrium. Let strategy vector $y$ be the one for which the two properties of the definition do not hold. Observe that the second property implies that $y$ is an equilibrium with respect to the players in the set $S = \{i | x_i \neq y_i\}$. This implies that the cost of any player $i \in S$ for which $y_i > 0$ is $h_i$ since he either plays a mixed strategy in an equilibrium or he plays the pure pay strategy. As the maximal cost a player can exhibit in an equilibrium is $h_i$, this implies that the only players in $S$ are ones for which $y_i = 0$. Now, by monotonicity of the play function, for every player $i \in S$ we have that $c_i(0, x_{-i}) \leq c_i(y) < c_i(x)$, in contradiction to the fact that $x$ is an equilibrium. □

**Corollary 4.6** *Every instance of the pay or play class admits at least a single semi-strong Nash equilibrium.*

This quite remarkable property that a semi-strong Nash equilibrium always exists ceases to hold once we remove the restriction that one of the strategies should have a fixed payoff. This is illustrated by the next example which is a variation on the prisoner's dilemma. For ease of exposition, the game is defined in terms of positive utility the players wish to maximize, instead of cost.

**Example 4.7** *Consider the following $3$-player game. Players $1$ and $2$ are paired together such that unless they pick the same strategy all the players have a utility of $0$. When players $1$ and $2$ choose the same strategy, the players utilities are defined by the following matrix where players $1$ and $2$ are the row player and player $3$ is the column player.*

|   | c   | d   |
|---|-----|-----|
| c | 4,4 | 0,0 |
| d | 6,0 | 1,1 |

*For brevity we only show that there is no mixed semi-strong Nash equilibrium. Let $p_1, p_2, p_3$ be the cooperation probabilities (strategy c) of the three players respectively. Then, player $1$ uses a mixed strategy if $4p_2 \cdot p_3 = (1 - p_2)(6p_3 + (1 - p_3))$. Similarly, player $2$ uses a mixed strategy if $4p_1 \cdot p_3 = (1-p_1)(6p_3+(1-p_3))$.*

*Therefore, we have that players $1$ and $2$ always play the same strategy, implying $p_1 = p_2$. Hence, player $3$ plays a mixed strategy if: $4p_1^2 = (1 - p_1)^2$.*

*By solving this system of equations we get that: $p_1 = p_2 = 1/3$ and $p_3 = 7/9$. To complete the proof, observe that this is not a semi-strong equilibrium since players $1$ and $2$ can deviate to the pure strategy d and increase their utility from $4/3 \cdot 7/9$ to $42/9$.*

## 5 Pareto Efficient Equilibria

One of the desirable properties of an equilibrium, increasing its stability, is Pareto efficiency. Roughly speaking, a strategy vector is Pareto efficient if any deviation that reduces the cost of one player (or more) strictly increases the cost of at least a single player. More formally:

**Definition 5.1** *An equilibrium $x$ is* Pareto efficient *if there is no strategy vector $y$, such that, for every player $i$, $c_i(y) \leq c_i(x)$, and for at least a single player the inequality is strict. If $y$ is restricted to be a pure strategy vector we say that $x$ is Pareto efficient with respect to pure deviations.*

We show that any Nash equilibrium of a "generic" pay or play game, i.e., a game in which players' best-responses are unique, is Pareto efficient. Formally, we define generic pay or play games as follows:

**Definition 5.2** *A pay or play game is* generic *if for every player $i$ and set of players $S$ such that $i \in S$: $h_i \neq g_i(S)$.*

We now prove the following:

**Theorem 5.3** *In a generic pay or play game, any pure Nash equilibrium is Pareto efficient with respect to pure deviations.*

**Proof:** Consider a Nash equilibrium $x$. Assume towards a contradiction that $x$ is not Pareto efficient. Let $y$ be a deviation reducing the cost of at least a single player. Define $S = \{i | x_i \neq y_i\}$. By the assumption that this is a generic game, we have that the cost of every player $i$ choosing the play strategy in $x$ is strictly less than $h_i$. Therefore, it has to be the case that for all players $j \in S$ it holds that $x_j = 1$. Now, similarly to our argument for the strong Nash equilibrium in Theorem 4.2, if there is a set of players that can reduce their cost by jointly switching from the pay strategy to the play strategy, then by monotonicity it is beneficial for a single player to perform this deviation. This is in contradiction to the fact that $x$ is a Nash equilibrium. □

**Corollary 5.4** *In a pay or play game, any pure Nash equilibrium in which every player $i$ who uses the play strategy incurs a cost strictly lower than $h_i$ is Pareto efficient with respect to pure deviations.*

Unfortunately, the previous theorem no longer holds for mixed deviations, as Example 4.3 illustrates.

Next, we demonstrate the importance of requiring the game to be generic. By tweaking the example from Claim 3.3 we create an instance in which in every equilibrium some players are indifferent between the two strategies, but their choice effects other players' cost.

**Claim 5.5** *The class of pay or play games contains games that possess pure Nash equilibria, and all such equilibria are not Pareto efficient.*

**Proof:** Consider the following game which includes four players numbered $0, 1, 2, 3$. The cost of player $i \in \{0, 1, 2\}$ is defined as: $h_i = 1.5$ for the pay strategy. $g_i(\{i-1, i, i+1\}) = 2$, $g_i(\{i-1, i\}) = 2$, $g_i(\{i, i+1\}) = 1.5$, $g_i(\{i\}) = 1$. Where $i+1$ and $i-1$ are computed modulo 3. The cost of player 3 is: $h_3 = 10$ and $g_3(S) = |S|$ for a set $S$ such that $3 \in S$. Observe that in all Nash equilibria exactly one player of the players $0, 1, 2$ chooses the pay strategy and the rest of the players choose the pay strategy. First, without loss of generality, we show that the strategy vector in which player 0 is the only one using the pay strategy is an equilibrium. Notice that player 1 is indifferent between the two strategies as both have a cost of 1.5. Players 2 and 3 strictly prefer the play strategy, hence this is an equilibrium. Next, we do a case by case analysis and show that any strategy vector in which the number of players using the pay strategy is not exactly one, is not an equilibrium.

1. There is no pure Nash equilibrium in which none of the players choose the pay strategy, since in this case one of the players $\{0, 1, 2\}$ can reduce its cost by choosing the pay strategy.
2. There is no pure Nash equilibrium in which two players (or more) choose the pay strategy. Clearly player 3 never choose the pay strategy. Now, if players $j$ and $j+1$ choose the *pay* strategy then if player $j$ switches to the pay strategy it reduces its cost to 1.

Observe that this equilibrium, in which a single player $i \in \{0, 1, 2\}$ chooses the pay strategy is not Pareto efficient. The reason is that, if player $i+1$ switches to the pay strategy then player 3 strictly benefit and the cost of the rest of the players remains the same. □

In the next section, we present in more depth one of the well studied games that belong to the pay or play class and show that every instance of this game admits a Pareto efficient pure Nash equilibrium.

## 6 Examples: Vaccination Games and Differential Pricing

The pay or play class is quite broad. In this section we focus on two well-studied subclasses of games that is contained in this class: vaccination games and differential pricing.

## 6.1 Vaccination Games

We first discuss the class of games presented by Aspnes et al. [1], which we refer to as "vaccination games". A vaccination game is played on a network $G$ with $|V| = n$ nodes that are the players of the game. Each player is faced with the following decision: buy a vaccination or not. If a player buys a vaccination then he pays a fixed cost, denoted by $c$. Else, the player risks getting his computer infected and exhibiting a loss of $l$. After all the players make their decisions one of the nodes in the network is selected uniformly at random to be infected by some virus. Next, the virus spreads in discrete rounds, such that in every round all the neighbors of every infected node that are not vaccinated get infected.

More formally, let $x$ be the strategy vector describing the decisions of the players whether to get the vaccine or not. $x_i = 1$ for a player that chooses to get the vaccine (pay) and $x_i = 0$ for a player that chooses not to get it (play). Denote by $R(x)$ the set of nodes choosing the pay strategy – getting the vaccine. Let $G_x$ be the attack graph that is constructed by removing all nodes in $R(x)$ and all their incident edges. The cost of the play strategy for node $i$ depends on the size of the connected component in $G_x$ that $i$ belongs to and the loss $l$. More precisely, the expected cost of the play strategy for a node $i$ in a connected component of size $k_i$ in $G_x$ is $\frac{k_i}{n} \cdot l$. It is not hard to see that this function is monotone increasing in the number of players choosing the play strategy and thus, this game belongs to the the pay or play framework.

It is shown in [1] that a pure Nash equilibrium for this game always exists. The proof is via a potential function, which relates the players' best responses to the size of the connected components in the attack graph. Let $\alpha = \frac{cn}{l}$. The set of pure Nash equilibria is characterized in [1] as follows: (1) every connected component of $G_x$ has a size of at most $\alpha$; and (2) for every player $i \in R(x)$ the size of its connected component in $G_x$ when node $i$ is added to the graph together with all its incident edges is at least $\alpha$.

By utilizing the framework of pay or play games, we are able to prove a new result for vaccination games – showing that a pareto-optimal Nash equilibrium always exists. As was discussed in the previous section, this property does not hold for pay or play games in general.

**Theorem 6.1** *The vaccination game admits a Pareto efficient Nash equilibrium.*

**Proof:** Assume without loss of generality that $l = 1$. This implies that the cost of the play strategy for player $i$ in strategy vector $x$ is simply the size of its connected component in $G_x + i$ divided by $n$. We refer to this as its infection probability. We show that there exists an equilibrium in which the infection probability of every node choosing the play strategy is strictly less than $c$. In other words, this implies that the size of every connected component of $G_x$ is strictly smaller than $c$. By Corollary 5.4 we have that this implies the equilibrium is pareto-optimal which completes the proof.

Assume towards a contradiction that in every equilibrium $x$ there exists a connected component of $G_x$ of size $c$. Let $x$ be an equilibrium for which $G_x$ has the minimal number of connected components of size $c$. Note that in case one of the connected components is not a tree, then it is possible to construct a new equilibrium with less connected components of size $c$. If the connected component is not a tree then there exists a node that can change its strategy to pay without harming the connectivity of its connected component in the attack graph. Denote this player by $i$. The new strategy vector is an equilibrium since player $i$ is indifferent between the two strategies. The only other affected players are ones in $i$'s connected component that still want to use the play strategy and ones using the pay strategy which are adjacent to $i$'s connected component. The adjacent nodes do not want to change their strategy to play since by doing that they will be a part of a connected component of size at least $c$, thus they do not want to switch.

Thus, it remains to handle the case in which all connected components of size $c$ are trees. Consider a leaf $i$ in one such tree, if this leaf is not connected to any other node (except its parent in the tree), then it can switch its strategy to play and it is still an equilibrium. Otherwise, it is connected to nodes who choose the pay strategy, denote this set of nodes by $S$. Go over the nodes in $S$ in some arbitrary order, for each node $j$ check the size of its connected component, if it is at most $c - 2$ change its strategy to play and continue. We claim that the resulting strategy vector is an equilibrium with a smaller number of connected components of size $c$. Observe that by construction the size of each connected component of the attack graph of the new strategy vector including neighbors of $i$, is smaller than $c$, therefore all nodes using the play strategy prefer it over the pay strategy. Also by construction, all the nodes in $S$ that use the pay strategy would be in a connected component of size at least $c$ if they decide to switch their strategy. Thus, the new strategy vector $y$ is an equilibrium such that $G_y$ has less connected connected components of size $c$ than $G_x$, a contradiction. □

## 6.2 Differential Pricing

Lastly, we briefly discuss another well-studied environment: differential pricing. Consider the following scenario: $n$ buyers are interested in purchasing some good, say a laptop. Each buyer has two options: (1) he can buy a laptop for a fixed price $p$ (there is a large enough supply of laptops to sell to all buyers); (2) take part in a lottery in which $k < n$ laptops will be assigned to $k$ bidders, uniformly at random, and each buyer who receives a laptop is charged a lower price $q < p$. (Of course, if there are less than $k$ buyers who decide to participate in the lottery, each of these buyers will be given a laptop). Observe that this can easily be formulated as a pay or play game. We note that every such environment admits a pure Nash equilibrium (and it is, in fact, a congestion game).

## 7 Price of Anarchy and Price of Stability

A natural metric for measuring the efficiency of a pure Nash equilibrium is by comparing its social cost (the sum of all players' costs) and the cost of the socially optimal solution (the strategy vector minimizing the sum of all players' costs). We present several simple results bounding the ratio between the optimal solution and worst pure Nash equilibrium (a.k.a price of anarchy) and the ratio between the optimal solution and best pure Nash equilibrium (a.k.a "price of stability") with respect to different restrictions on the cost functions. We begin with a positive result showing that for a very restricted subclass of pay or play games the price of anarchy is 2:

**Claim 7.1** *If all players have the same submodular cost functions, and the cost function does not depend on players' identities, then the (pure) price of anarchy is bounded by 2.*

**Proof:** Consider a specific pure Nash equilibrium $x$ and optimal solution $o$. Denote by $k_x$ and $k_o$ the number of players using the play strategy in $x$ and $o$ receptively. Observe that if $k_o \leq \frac{n}{2}$, then at least $n/2$ players choose to pay and hence the cost of the optimal solution is at least $\frac{n}{2} \cdot h$. The cost of the Nash equilibrium is at most $n \cdot h$, since players can always choose the pay strategy and pay $h$. Thus, the price of anarchy for this case is at most 2.

We are left with the case that $k_o > \frac{n}{2}$. Observe that this trivially implies that $k_x \leq 2k_o$. Also, it is not hard to see that $k_o \leq k_x$. Now, consider the difference in cost between the pure Nash equilibrium and the optimal solution: $c(x) - c(o) = ((n - k_x)h + k_x \cdot g(k_x)) - ((n - k_o)h + k_o \cdot g(k_o)) = (k_o - k_x)h + k_x \cdot g(k_x) - k_o \cdot g(k_o)$. The fact that $k_n < \frac{k_o}{2}$, together with the submodularity of the cost function, and the fact that the cost function is nondecreasing, imply that $g(k_x) \leq g(2k_o) \leq 2g(k_o)$. Hence, $c(x) - c(o) < (k_o - k_x)h + 2k_x \cdot g(k_o) - k_o g(k_o) = (k_o - k_x)h + (2k_x - k_o)g(k_o) \leq k_x \cdot g(k_o) \leq n \cdot g(k_o) \leq c(o)$, where the last two inequalities follow from the simple observation that $h \geq g(k_o)$ □

Next, we show that once we lift either of the two restrictions previously imposed: (1) all players have the same cost functions, (2) the cost function depends only on the number of players choosing the play strategy, the price of stability can be very high:

**Claim 7.2** *The (pure) price of stability of a game with player-specific cost functions that are not dependent on players' identities can be linear in $n$.*

**Proof:** Consider the following $n$-player instance, for player 1, $h_1 = n + \epsilon$ and $g_1(S) = |S|$ for $i \in S \subseteq N$. For player $j \neq 1$ $h_j = 2\epsilon$ and $g_j(S) = \epsilon$ for $j \in S \subseteq N$. In the optimal solution of this instance, player 1 is the only one choosing the play strategy – the cost is $1 + 2(n-1)\epsilon$. On the other hand, in the unique Nash equilibrium all players choose the play strategy, the social cost in this case $n + (n-1)\epsilon$. □

**Claim 7.3** *If all players have the same submodular cost function (possibly depends on the players' identities) then the (pure) PoS can be linear in $n$.*

**Proof:** Consider the following instance where $h = 1 + \epsilon$ and for any set $S$ such that $1 \notin S$ we define $g(\{1\} \cup S) = 1$ and $g(S) = 0$. Then, in the optimal solution player 1 chooses the pay strategy, for a social cost of $1 + \epsilon$. In any Nash equilibrium all players choose the play strategy for a total cost of $n$. □

## 8 Conclusions

We introduced the pay or play framework, which captures a simple scenario in which decision makers select between certain and uncertain outcomes, and the realization of the uncertain outcome is solely dependent on the decision makers and not on "nature". We studied the properties of equilibria (existence, efficiency, complexity, and more) in pay or play games from both a game-theoretic perspective and a computational perspective. Our main positive result established that games in this class always possess a semi-strong equilibrium. We regard our results for pay or play as a first step, and believe that further exploring the game-theoretic and computational properties of this class of games (and its subclasses) can provide valuable insights into strategic decision making under uncertainty.